\documentclass[11pt]{article}

\usepackage{amsthm}
\usepackage{amssymb}
\usepackage{amsmath}

\DeclareMathAlphabet\EuFrak{U}{euf}{m}{n}	
\SetMathAlphabet\EuFrak{bold}{U}{euf}{b}{n}	

\begin{document}
\author{Sergio Doplicher 
                         \\Dipartimento di Matematica
                         \\University of Rome "La Sapienza" 
                         \\00185 Roma, Italy  }

\title{Spacetime and Fields, a Quantum Texture} 
\maketitle

\begin{abstract} 
We report on joint works, past and in progress, with
K.Fredenhagen and with J.E.Roberts, on the quantum structure of spacetime
in the small which is dictated by the principles of Quantum Mechanics and
of General Relativity; we comment on how these principles point to a
deep link between coordinates and fields. This is an expanded version of
a lecture delivered at the 37th Karpacz School in Theoretical Physics,
February 2001.
\end{abstract}

\section{Spacetime Uncertainty Relations}
At large scales spacetime is a pseudo Riemanniann manifold locally modeled
on Minkowski space. But the concurrence of the principles of Quantum
Mechanics and of Classical General Relativity points at difficulties at
the small scales, which make that picture untenable. For if we try to
locate an event in say a spherically symmetric way around the origin in
space with accuracy $a$, according to Heisenberg principle an
uncontrollable
energy $E$ of order $1/a$ has to be transferred, which will
generate a gravitational field with Schwarzschild radius $R \simeq E$ ($
\hbar = c =  G  = 1$). Hence $a \gtrsim R \simeq  1/a$  and $a
\gtrsim 1$, i.e. in CGS units

\begin{equation}
\label{dopleq1}
a \gtrsim \lambda_P \simeq 1.6 \cdot 10^{-33} cm. 
\end{equation}

If however we measure one of the space coordinates of our event with great
precision $a$, but allow large uncertainties $L$ in the knowledge of the
other coordinates, the energy $1/a$ may spread over a thin disk of radius L
and thus generate a gravitational potential that would vanish everywhere 
as $L \rightarrow \infty$. 

One has therefore to expect Space Time Uncertainty Relations emerging from
first principles, already at a semiclassical level. Carrying through such
an analysis \cite{dopl1,dopl2} one finds indeed that, if the smallest and
largest space uncertainties of an event are denoted by $a,b$ respectively,
and the time uncertainty by $\tau$  , the gravitational potential
generated by the energy $1/ \min(a,\tau )$ localized at some instant with
accuracies $a,b,\tau$, is at most of the order 

\begin{equation}
\label{dopleq2}
\left| V \right|  \simeq  \frac1 {b \cdot \min(a,\tau)}
\end{equation}

Now our basic requirement is that the localization experiment should not
deform spacetime in such a way that no signal from the region we wish to 
observe can reach infinity in space, otherwise this would put the observed
event out of reach for any distant observer; namely

\begin{equation}
\label{dopleq3}
g_{00}  =    1  +  2V   >    0,
\end{equation}

where $V$ is the potential generated by the energy transferred with the
localization measurement itself; hence by (\ref{dopleq2}) a necessary 
condition is

\begin{equation}
\label{dopleq4}
       b \cdot \min(a,\tau  )  \gtrsim   1.
\end{equation}

The Space Time Uncertainty Relations strongly suggest that spacetime has a
Quantum Structure at small scales, expressed, in generic units, by

\begin{equation}
\label{dopleq5}
       [q_\mu  ,q_\nu  ]   =   i \lambda_P^2   Q_{\mu \nu}
\end{equation}

where $Q$ has to be chosen not as a random toy mathematical model, but in
such a way that necessary restrictions like (\ref{dopleq4}) 
follow from (\ref{dopleq5}). Further we
want
to impose (full) Lorentz invariant conditions on $Q_{\mu \nu}$ , so that
our models
are compatible with Special Relativity; since in (\ref{dopleq5}) $Q$ is 
dimensionless, the commutator will effectively vanish for large distances
compared to the Planck scale.

But we do not insist on covariance under general coordinate transformations, 
which, at a quantum level and at small scales, cannot be supported by
conceptual experiments, as the freely falling laboratory, in presence of
fields which vary significantly over Planckian distances. Moreover, for
the sake of Elementary Particle Physics, an asymptotically flat background
is an appropriate idealization, for the distribution of masses in the
Universe should not affect significantly the outcome of collision
experiments in our laboratories.

The noncommutativity of the operators $q_0  ,...,q_3$ can be measured by
the
fundamental invariants

\begin{equation}
\label{dopleq6}\left. 
\begin{array}{l}
Q_{\mu \nu} Q^{\mu \nu} ; \\
\left[q_0  ,...,q_3 \right] := \det \left( 
\begin{array}{ccc}
q_0 & \cdots  & q_3 \\
\vdots  & \ddots  & \vdots  \\ 
q_0 & \cdots  & q_3
\end{array}
\right) := \\ {} \\
\varepsilon^{\mu \nu \lambda \rho} q_\mu q_\nu q_\lambda q_\rho =  - (1/2) Q_{\mu \nu}  (*Q)^{\mu \nu}
\end{array}
\right. 
\end{equation}

but the second is invariant only under the proper Lorentz transformations
and only its square is invariant under space and time reflections as well. 

If we (temporarily) assume that the components of $Q$ commute with one
another, and let ${\bf e}, {\bf m}$ denote the triples of electric
respectively
magnetic
components, we have

\begin{equation}
\label{dopleq7}
(- 1/2) Q_{\mu \nu}   Q^{\mu \nu}    =    {\bf e}^2   - {\bf m}^2  ;
\end{equation}

since ${\bf e}$ and ${\bf m}$ respectively govern the space-time and
space-space
uncertainty relations, symmetry and (\ref{dopleq4}) suggest the condition

\begin{equation}
\label{dopleq8}
 Q_{\mu \nu}   Q^{\mu \nu}    =    0.
\end{equation}

Therefore the basic Quantum Condition must read

\begin{equation}
\label{dopleq9}
[q_0  ,...,q_3  ]^2      =     S,
\end{equation}

where $S$ is a Lorentz invariant.  

We will see later how more general choices for $S$ are important, but
(\ref{dopleq4})
suggest a multiple of $I$. If we also require that the $Q$ commute with 
the $q$, we get the Basic Model introduced and discussed in detail in
\cite{dopl1} that we will briefly report on in the next Section.

Other approaches to uncertainty relations affected by gravity and related
phenomena can be found e.g. in \cite{dopl7},..., \cite{dopl15}. We do not
attempt
to give a
complete list of references related to this subject, which became quite
numerous in the last three years; approaches based on the quantum
deformations of the Poincar\'e Algebra received a lot of attention, cf
\cite{dopl17} and references therein.

\section{The Basic Model}

In the notation introduced above the quantum conditions of the Basic
Model may be rewritten as

\begin{equation}
\label{dopleq10}
\left[ q_\mu  , Q_{\lambda \rho}  \right]  =  0,  
\end{equation}
                
\begin{equation}
\label{dopleq11}
 {\bf e}^2     =   {\bf m}^2  ,    {\bf e} \cdot  {\bf m}  = \pm I.
\end{equation}

In this model the following weaker form of (\ref{dopleq4}) is implemented,
cf \cite{dopl1,dopl2} :

\begin{equation}
\label{dopleq12}
\Delta q_0 \cdot \sum \limits_{j = 1}^3 \Delta q_j \gtrsim 1 ; \sum
\limits_{1 \leq j < k \leq 3 } \Delta q_j  \Delta q_k \gtrsim 1 .
\end{equation}

Relations (\ref{dopleq11}) define an algebraic manifold with two connected
components
each isomorphic to the coset space of the proper Lorentz group modulo
boosts along a fixed direction and rotations around it, i.e. to
$SL(2,\mathbb C) / {\mathbb C}_*$,
where $\mathbb C_*$ is embedded in $SL(2,\mathbb C)$ as the 1,1 component
of diagonal matrices.
Each pair $({\mathbf e}',{\mathbf m}')$ as in (\ref{dopleq11}) can be obtained from
a pair such that ${\bf e} = \pm {\bf m}$ 
by a boost with velocity, say  ${\bf v}$, orthogonal to ${\bf e}$, hence
by (\ref{dopleq11}) ${\bf e}$ and
 ${\bf m}$ are vectors in the unit sphere $S^2$   in three dimensional
space, and $ {\bf v}$ is
a tangent vector to $S^2$   ; summarizing

\begin{equation}
\label{dopleq13}
    \Sigma_+ \simeq \Sigma_-  \simeq   SL(2,\mathbb C)/ {\mathbb C}_*
\simeq TS^2 .
\end{equation}

While classical Spacetime is described by the commutative C* algebra of 
continuous functions vanishing at infinity, Quantum Spacetime will be 
described by a noncommutative C* algebra $\mathcal E$, to which the $q$
are 
affiliated in the sense of \cite{dopl18}, cf \cite{dopl1},  i.e  each
representation of $\mathcal E$
determines  operators $q_\mu$    fulfilling our Quantum condition, and all
"regular" 
representations appear this way. 

We adopted in  \cite{dopl1}  the following paradigm, which may well apply
to
more general cases (cf \cite{dopl4}): if we interpret (\ref{dopleq5}) as
defining a bundle of
Lie algebras, in that case over $\Sigma$, the regular
representations will 
be those which are integrable to a representation of the corresponding 
bundle of simply connected Lie groups; the C* algebra  $\mathcal E$
will then 
arise as a continuous field of group C* algebras. 

In the basic model the fibers are just Heisenberg groups with a non 
degenerate $\mathbb C$-number commutator matrix (the generic point  in
$\Sigma$ 
), 
hence we get a continuous  field of the algebra of all compact operators 
(on a separable infinite Hilbert space) which can be proved to be trivial 
(see \cite{dopl1}), i.e. 

\begin{equation}
\label{dopleq14}
\mathcal E \simeq {\mathcal C}_0 (\Sigma , \mathcal K) .
\end{equation}

These findings fit very well in the theory of strict deformation
quantization \cite{dopl19}.

This C* algebra carries a natural action of the (full) Poincar\'e group
$\mathcal P$
as
automorphisms, which is actually determined by its extension to the
affiliated $q$'s, fulfilling the natural relations

\begin{equation}
\label{dopleq15}
\alpha_L (q)    =    L^{-1} q \quad , \quad   L \in \mathcal P   .
\end{equation}

Thus  $\mathcal E$     is a Quantum space but its global symmetries are
the classical
ones, as expected since at large scales the model turns classical again,
and the Poincar\'e transformations are global motions, acting the same way
in the small and in the large. This situation parallels the familiar one
in nonrelativistic Quantum Mechanics, where the Schroedinger Operators $q$
and $p$ do not commute, but the Galilei invariance is expressed by an
action
of the classical Galilei Group as automorphisms (representations up to a
phase appear only in the unitary implementations). Actually this structure
is indeed a special case of our present model, cf below.

The classical concept of points in a space has to be replaced by pure
states with minimal uncertainties, i.e. pure states which are optimally
localized in the sense that the quantity

\begin{equation}
\label{dopleq16}
(\Delta q_0)^2 + ... + (\Delta q_3)^2
\end{equation}

is minimal; this is a frame dependent condition, which picks a point ${\bf 
e} = \pm {\bf m}$ in the spectrum of the $Q$'s, i.e. a point in the base 
$S^2\times \left\{ \pm 1 \right\}$ if we
think of  $\Sigma$  as a tangent manifold, so that the $q$'s fulfilling
(\ref{dopleq5}) now
become the Schroedinger operators $q,p$ for a particle in two dimensions
(the four dimensional translations acting as Galilei transformations),
and the expression (\ref{dopleq16}) being minimal implies that its value 
is $2$ and that our
state is the
ground state of the harmonic oscillator for those Schroedinger operators.

Such states ought to have a preferred role in discussing the large scale
limit of the Quantum space; since in (\ref{dopleq14}) the Planck length
appears only
in the exponential in the Weyl relations, which force the fiber to be
$\mathcal K$,
and in the large scale limit $\mathcal K$ deforms to ${\mathcal C}_0  
({\mathbb R}^4 )$, we see that the
quantum spacetime becomes  ${\mathbb R}^4  \times \Sigma$ in the
large scale limit, while,
if only optimally localized states are considered, the limit is rather

$$
{\mathbb R}^4   \times  S^2   \times  \left\{ \pm 1 \right\} .
$$

Thus the discrete space $\left\{ \pm 1 \right\}$ appears because the
spectrum of the centre
of the algebra generated by the $Q$'s is not connected (a consequence of
imposing symmetry under reflections too), and while the continuous factor 
in the ghost manifold is not compact, only a compact manifold, actually a 
sphere with radius the square of the Planck length, plays a role if we are 
testing with optimally localized states.

The paradigm we adopted in attaching a C* algebra to relations
(\ref{dopleq5}) in our
model tells us how to calculate functions $f(q)$: as in the von Neumann
-Wigner-Moyal calculus, write $f$ as the Fourier transform of its
ordinary anti-Fourier transform, and replace the exponentials by the Weyl
operators $\exp i ( \alpha q)$; the multiplication of these exponentials
is
precisely governed by the bundle of Lie groups associated to the models;
thus this paradigm can, and will be, applied in some more general context.
Moreover space integration at time t and spacetime integration can be
easily defined and related to the trace in each fiber, so that we can
introduce Free Fields on QST, the free Hamiltonian, which turns out to be
unchanged by the quantum deformation, and interaction Hamiltonians,
i.e. we can lay down the setup to apply the usual perturbation expansion
(cf \cite{dopl1}). 

While integration over space or spacetime poses no problem in this
model,
integration over   $\Sigma$ does, since we tacitly assumed that our
fields
do
not depend on the points of $\Sigma$        ; but we have no
bounded invariant
measure on   $\Sigma$     so we cannot integrate to get an invariant
result. 

The way out chosen in   \cite{dopl1} was to integrate over the base $S^2
\times \left\{ \pm 1 \right\}$ of $\Sigma$, 
thus keeping only rotation invariance; but in the end we face a more
serious difficulty. Namely the perturbation expansion is found to be 
exactly that of a non local theory on the classical Minkowski space.

Of course  (\ref{dopleq4}) suggests that causality breaks down at short
distances: but
it should be recovered at large scales with respect to Planck length (say
at QCD scales, $10^{-17}$ cm.), while the acausal effects of ordinary
nonlocal theories might cumulate after summing the perturbation expansion
in a disruptive way.

Strangely enough, these lessons of \cite{dopl1} have been largely
neglected; well
after the appearance of \cite{dopl1} we assisted to a flow of papers on
QFT models on a QST which is characterized by (\ref{dopleq5}) with a
fixed $\mathbb C$-number
tensor on the right, disregarding the physical meaning of
noncommutativity and the need of Lorentz invariance, but extensively 
applying the calculation aspect of what we exposed, summarized by the use
of the ``star product''.

The negative conclusions referred to above might lead us to reanalyze the
concept of interaction over QST; in particular the ordinary Wick product
should not be allowed: if e.g. we are to define the Wick square of the
field $A$, we can evaluate $A(q)A(q')$ on distinct variables $q$, $q'$,
but then
we cannot set $q - q' = 0$, since these operators obey similar relations
to
(\ref{dopleq4}); we can however evaluate a conditional expectation
defined by an
optimally localized state on $q - q'$; as this is the ground state of an
harmonic oscillator, it introduces a Gaussian nonlocality factor which
violates again causality, but might fully regularize the theory, and in
fact might give rise to a Gaussian fall off of cross sections at large
energies  \cite{dopl5}.

The problems with causality lead us \cite{dopl5}  to enquire about light
propagation, i.e. classical ED on QST; while the local gauge group of
Classical ED on Minkowski space is the unitary group of ${\mathcal C}_0 ( 
{\mathbb R}^4 ) + {\mathbb C} \cdot I$,
it is that of $\mathcal E + {\mathbb C} \cdot I$ in the case of QST.
Treating it the usual way
we found that ED is characterized by nonlinearly selfinteracting
equations, for which a plane wave is a stationary solution, but
superpositions of two different plane waves are not, with a propagation
into massive modes; in principle, this effect ought to be detectable 
splitting a monochromatic laser beam into a superposition of states with 
different momenta with the help of a partially reflecting mirror (or detecting 
the light of a distant galaxy split by a gravitational lens); but the fraction 
of energy density which would go into these massive modes, calculated to the 
lowest order in the Planck length, turns out to be of order 
lower than $10^{-130}$.

More seriously, such a theory has a huge gauge group, so it is difficult to
propose testable effects, no matter how tenuous, in terms of gauge invariant
quantities.

For recent discussions of possible testable effects of Quantum Gravity cf
e.g. \cite{dopl21,dopl22}.

Another drastic consequence is the nonvanishing of the current divergence,
due to quantum gravitational anomalies.

But the Gauge Principle expresses the point nature of interactions, and is
the basic principle lying behind locality in ordinary QFT,
so it might well by itself provide a rigid substitute to causality in QFT
on QST. This hope motivates a long standing attempt to a general
formulation of gauge theories on a noncommutative manifolds using the
absolute differential calculus (\cite{dopl5,dopl6}; this calculus emerged
also
in other papers appeared meanwhile, cf e.g. \cite{dopl16}). One might
expect
that a proper noncommutative approach might lead to a new picture of
interactions at Planck scale, which avoids the unpleasant features met 
when we just replace products with $*$-products.

An approach to gauge theories on noncommutative spaces based on the notion
of covariant coordinates has been proposed in \cite{dopl24} and references
therein.

\section{Deformed Models}
The remaining sections are based on work in progress with K.\ Fredenhagen,
D.\ Bahns and G.\ Piacitelli.

The basic model discussed in the previous section has many virtues
including simplicity, even if it does not implement in full relations
(\ref{dopleq4}). However these relations appear, as relations
(\ref{dopleq12}), only necessary to guarantee (at least at a semiclassical
level) the
gravitational stability of localization of events, but not a
priori sufficient to that purpose; furthermore it might well turn out to
be impossible to formulate a necessary and sufficient condition which
involves solely the  background geometry, without a dynamical description
of spacetime, cf next section. 
However, if we relax the drastic simplification (\ref{dopleq10}) that the
$Q$'s are
central, and instead 
we only assume that they commute with one another, keeping the other
Quantum Conditions, there is room for deformed models where the Spacetime
Uncertainty relations are implemented in forms stronger than 
(\ref{dopleq12}).

One such model (\cite{dopl5}; partly announced in \cite{dopl4}) can be
formulated
introducing self adjoint central operators,  which form  two antisymmetric
2-tensors $H,T$ and a four vector $C$, and adding to the $q$'s two scalar
commuting generators $R,S$, and imposing

\begin{equation}
\label{dopleq17}\left. 
\begin{array}{l}
\left[ q_\mu   ,q_\nu \right]   =   i( H_{\mu \nu} + T_{\mu \nu} R), \\ 
{} \\
\left[ q_\mu   ,R \right]     =   i C_\mu S , \\ 
{} \\
\left[ q_\mu   ,S \right]     =  -i C_\mu R , \\
{} \\
\left[ R,S \right]        =   0  ,
\end{array}
\right. 
\end{equation}

where $S^2 + R^2 $ is central and can be set equal to $I$,  and, by
Jakobi identity,

\begin{equation} 
\label{dopleq18}
C_\mu   T_{\nu \lambda}   +   C_\nu   T_{\lambda \mu}   +   C_{\lambda}
T_{\mu \nu } = 0 , \quad for \quad all \quad \mu , \nu , \lambda .
\end{equation}

Furthermore, the contraction of $T$ and of its Hodge dual $*T$ with itself
and
with $H$ should vanish, while $H$ fulfills the same conditions as $Q$ in
the
basic model.

The full Lorentz group $\mathcal L$ will act transitively on the spectrum
of the centre, so that here, at large scales, the classical limit of our
QST will be

\begin{equation}
\label{dopleq19}
{\mathbb R}^4    \times  \mathcal L
\end{equation}

a manifold with $10$ dimensions, which would effectively reduce here
too if we restricted attention to optimally localized states.
A more detailed account of this model will be discussed elsewhere \cite
{dopl27}.

\section{A Dynamical Picture of Quantum Spacetime}
The models of QST outlined above try to implement in the noncommutative
nature of the underlying geometry some of the minimal limitations on the
localization of an event which are imposed by our present knowledge of the
principles of Physics. Developing QFT in the appropriate way on this
underlying geometry rather than on Minkowski space should avoid some of
the contradictions we would be otherwise bound to meet.  But we might
expect that the very structure of spacetime in the small, hence the
algebraic structure of the underlying model, should depend on the
dynamics, and thus on the quantum state. We propose a general scenario 
where this would be the case. 

Let us first note that if we develop QFT over a fixed geometric background
described by a (noncommutative C*) algebra $\mathcal E$,  carrying an
action of
the Poincar\'e group by automorphisms, the (bounded functions of the)
field
operators (or more generally \cite{dopl23} local observables) should take
values in a quasilocal C* algebra $\EuFrak A$, and fields would be
(functions from QST to $\EuFrak A$) described by elements of (or affiliated to) the tensor
product

\begin{equation}
\label{dopleq20}
\mathcal E \otimes \EuFrak A
\end{equation}

But a more realistic picture of QST might well involve operators in
(\ref{dopleq20}) which cannot be easily split in the two factors; the
commutators of the $q$'s would then appear as functions of the fields,
more specifically of
the metric  $g_{\mu \nu}$, coupled to all fields by Einstein
Equation, the fields
themselves being at the same time functions of the $q$'s. Thus the
commutation relations between the $q$'s should appear as part of the
equations of motion:

\begin{equation}
\label{dopleq21}\left. 
\begin{array}{l}
\left[ q_\mu  ,q_\nu  \right]       =  i Q_{\mu \nu}   (g) \\ 
{} \\
R_{\mu \nu}  - (1/2)Rg_{\mu \nu}   =  8 \pi  T_{\mu \nu}  (\psi) \\ 
{} \\
F(\psi ) =  0,
\end{array}
\right. 
\end{equation}

where $T_{\mu \nu}( \psi )$  is the energy momentum tensor of the fields
involved,
except gravitation, and the last line is symbolic for their equation of
motion, where $g$ enters too through the covariant derivatives. Of course
the action of translations on the $q$'s will no longer 
be just the addition of multiples of the identity, since the $q$'s depend 
on the metric $g$ on which translations act as well.

As an attempt to investigate the form of (\ref{dopleq21},a), suppose we
adopt a
semiclassical approach, replacing the right hand side of
(\ref{dopleq21},b) by its
expectation value in a given state and let g be a classical solution;  
if we perform a measurement to localize an event in this state, we should
repeat the considerations of \cite{dopl1}, cf Section 1 above, in the
background $g$;
in the approximation of linearized gravity, with $V$ as in equation
(\ref{dopleq3}), we
should now impose

\begin{equation}
\label{dopleq22}
g_{00} + 2V  >   0;
\end{equation}

hence

\begin{equation}
\label{dopleq23}
g_{00} \cdot b \cdot \min(a,\tau  ) \gtrsim   1.
\end{equation}

If we forget for a moment not only general covariance, according to which
g should have no intrinsic meaning, but even Lorentz covariance, we could 
fulfill (\ref{dopleq23}) requiring

\begin{equation}
\label{dopleq24}
\left[ q_\mu   , q_\nu  \right]  =  i Q_{\mu \nu}  g_{00}^{-1}
\end{equation}

where $Q_{\mu \nu}$ does not depend of $g$, and is defined as above in
this report.

According to General Relativity the Ricci tensor $R_{\mu \nu}$ is
physically
significant but the metric tensor $g$ is not; yet it has been proposed
\cite{dopl15}
that Quantum Mechanics might alter this view, a possibility to be kept in
mind while trying to rewrite a more convincing covariant extrapolation of
(\ref{dopleq24}).  

The first natural guess would be to replace $g_{00}^{-1}$ in
(\ref{dopleq24}) by a scalar depending only on the local variations of
$g$,  as the scalar curvature $R$; hence, using Einstein Equation
(\ref{dopleq21},b) we would write

\begin{equation}
\label{dopleq25}
[  q_\mu   , q_\nu  ]  = - 8 \pi \alpha  i Q_{\mu \nu}  g^{\lambda \rho}
T_{\lambda \rho} (\psi ),
\end{equation}

where a further constant factor $\alpha$ has been allowed; or, even more
generally, we could replace the Quantum Conditions by

\begin{equation}
\label{dopleq26}\left. 
\begin{array}{l}
Q_{\mu \nu}   Q^{\mu \nu} = 0, \\
{} \\
\left[ q_0  ,...,q_3  \right]^2  = (\alpha R)^4  .
\end{array}
\right. 
\end{equation}

Equations (\ref{dopleq25}) and (\ref{dopleq26}) do not reduce to our
background model where $R = 0$,
so we are tempted to replace $\alpha R$ by $I + \alpha  R$; we limit
ourself here to
support the scenario expressed in (\ref{dopleq21}) without committing
ourself to a
choice, but point out (maybe only as a curiosity) that if our state is
strictly localized in a tiny region, expectations of observables which are
spacelike to that tiny region will be the same as in the vacuum and there we 
will find the following semiclassical approximation to (\ref{dopleq25})

\begin{equation}
\label{dopleq27}
[ q_\mu   , q_\nu   ]   =  - 8 \pi \alpha  i Q_{\mu \nu}    g^{\lambda
\rho} \langle T_{\lambda \rho}\rangle_0 ;
\end{equation}

we might here insert the empiric evidence that $\langle T_{00} \rangle_0$ is not zero
but
equal to the cosmological constant $\Lambda$   ; in a relativistic vacuum

$$
\langle T_{\lambda \rho} \rangle_0    = \Lambda \cdot  diag (1, -1, -1, -1);
$$

now if $g$ is a spherically symmetric stationary solution with $g_{j 0}   =
g_{0j} = 0 ,\quad j = 1,2,3$, and $- {\bf g}$ is its space part,
(\ref{dopleq27}) takes the form

$$
[ q_\mu   , q_\nu   ]   =   - 8 \pi \alpha \Lambda i Q_{\mu \nu}   ( g_{00}^{-1}
+ tr({\bf g}^{-1} ));
$$

for the Schwarzschild solution, for instance, the last term in brackets would be
equal to   $g_{00}^{-1}   +   g_{00}   +  2$; but if there is a preferred
frame (that of the 
Cosmic Background Radiation) where  $\langle T_{00}\rangle_0=\Lambda$, 
$\langle T_{jj}\rangle_0 
=0$, $j =1,2,3$, we would get exactly (\ref{dopleq24}).  

These comments do not pretend to be neither satisfactory nor in a final shape
(we used in our heuristic argument strict locality, which is bound to
fail at Planck distances); yet it might well turn out that the quantum
nature of spacetime does say  something on the problem of the cosmological
constant; for the presence of T in the right hand side of our spacetime 
commutation relations should imply an effective repulsion at short
distances, and since quantum spacetime links aspects in the small
(ultraviolet) to aspects in the large (infrared), this  short range
repulsion might well give rise to long range effects.

\section{Hints of Relations to String Theory}
With the notation of Section 1, our Space Time Uncertainty Relations read

\begin{equation}
\label{dopleq28}\left. 
\begin{array}{l}
a \cdot b  \gtrsim  1, \\ 
{} \\
\tau \cdot  b  \gtrsim  1;
\end{array}
\right. 
\end{equation}

the second one had been actually proposed earlier on the basis of a qualitative
argument in String Theory \cite{dopl7}, and derived later in the context
of D-branes \cite{dopl11,dopl8}. Other recent findings in that domain lead
to relations similar to our first relation too \cite{dopl16}. 

Other superficial coincidences can be noted: $U(1)$ gauge theory on QST
described
by the basic model is actually a $U( \infty  )$ gauge theory (more
precisely, the
gauge group will be the unitary group of  $ \mathcal E  +  \mathbb C \cdot 
I$,
namely the product of the torus  $\mathbb T$ with the group of continuous 
functions equal to $I$ at infinity from $\Sigma$ to the group of
unitaries which are perturbations of $I$ by a compact operator), while
$U(N)$ gauge theory in the
limit $N \rightarrow \infty$ is believed to merge with String
Theory; the QST version of 
Wick product leads to ultraviolet finite theories with a Gaussian like
falloff of the transition matrix elements above Planckian energy-momentum 
values \cite{dopl26}.

These facts might be no more than fortuitous coincidences, but suggest that the
physical principles underlying the proposal of Quantum Spacetime might even turn
out to provide the fundamental physical motivations which are still lacking in
String Theory.


\begin{thebibliography}{Enquat 23}

\bibitem{dopl1}
S.Doplicher, K.Fredenhagen, J.E.Roberts: {\em The Quantum
Structure of Spacetime at the Planck Scale and Quantum Fields}, Commun.
Math. Phys. 172, 187 - 220 (1995);

\bibitem{dopl2}
S.Doplicher, K.Fredenhagen, J.E.Roberts:  {\em Spacetime
Quantization Induced by Classical Gravity}, Phys. Letters B 331, 39 - 44
(1994);

\bibitem{dopl3} 
S.Doplicher:  {\em Quantum Physics, Classical Gravity, and Noncommutative
    Spacetime}, Proceedings of the XIth International Conference of Mathematical
    Physics, D.Iagolnitzer ed, 324 - 329, World Sci. 1995;

\bibitem{dopl4} 
S.Doplicher: {\em Quantum Spacetime}, Annales Inst. Henri
Poincare' vol.64, 543 - 553, 1996;

\bibitem{dopl5} 
work in progress with D.Bahns, K.Fredenhagen and G.Piacitelli;

\bibitem{dopl6} 
K.Fredenhagen: address to the Goslar Meeting, 1998; {\em Quantum
fields and noncommutative space-time}, Proceedings of the Hesselberg
Meeting March 1999, F.Scheck, W.Werner and H.Upmeier eds., Springer
L.N.P. 596 (2002);

\bibitem{dopl7} 
T.Yoneya: {\em Duality and Indeterminacy Principle in String
Theory} , in {\em Wandering in the Fields}, K.Kawarabayashi,A.Uwaka eds., World Sc. (1987);

\bibitem{dopl8} 
T.Yoneya: {\em String Theory and Space-Time Uncertainty Principle}, 
hep-th/0004074;

\bibitem{dopl9} 
C.A.Mead: {\em Possible Connection between Gravitation and Fundamental
Length}, Phys. Rev. 135B, 849-862 (1964);

\bibitem{dopl10} 
D.Amati, M.Ciafaloni, G.Veneziano: {\em Can Spacetime be probed
below the String Size?} Phys. Lett. B, 216 41 (1989);

\bibitem{dopl11} 
G.Amelino-Camelia, J.Ellis,N.E.Mavromatos,D.V.Nanopoulos: {\em On
the Spacetime Uncertainty Relations of Liouville Strings and D-Branes}, 
hep-th/9701144;

\bibitem{dopl12} 
J.Lukierski, A.Nowicki, H.Ruegg: {\em New Quantum Poincare'
Algebra and k-deformed field theory}, Phys. Lett.B 293, 344-352 (1992);

\bibitem{dopl13} 
A.Kempf: {\em Uncertainty relations in quantum Mechanics with
quantum group symmetries}, J. Math. Phys. 35, 4483-4496 (1994);

\bibitem{dopl14} 
M.Maggiore: {\em Quantum groups, gravity and the generalized uncertainty 
principle}, Phys. Rev. D49, 5182-5187 (1994);

\bibitem{dopl15} 
Chong-Sun-Chu, Pei-Ming Ho, Yeong-Chuan Kao: {\em Worldvolume
uncertainty relations for D-branes}, hep-th/9904133;

\bibitem{dopl16} 
S.Cho, R.Hinterding, J.Madore, H.Steinacker: {\em Finite Field
Theory on Noncommutative Geometry }, Int. J. Mod. Phys. D9, 161-199
(2000);

\bibitem{dopl17} 
P.Kosinski, J.Lukierski, P.Maslanka: {\em Noncommutative
parameters of quantum symmetries and Star Products} hep-th/0012056;

\bibitem{dopl18} 
S.L.Woronowicz: {\em Unbounded Elements Affiliated with C*
Algebras and non compact Quantum Groups}, Commun. Math. Phys. 136, 399-432
(1991);

\bibitem{dopl19} 
M.A.Rieffel: {\em On the Operator Algebra for the Spacetime
Uncertainty Relations}, in {\em Operator Algebras and Quantum Field Theory},
S.Doplicher, R.Longo, J.E.Roberts and L.Zsido eds, I.P. 1997;

\bibitem{dopl20} 
D.V.Ahluwalia, {\em Principle of equivalence and wave-particle
duality in quantum gravity}, gr-qc/0009033; 

\bibitem{dopl21} 
G.Amelino-Camelia: {\em Gravity mediated interferometers as
probes of a low-energy effective quantum gravity} gr-qc/9903080;

\bibitem{dopl22} 
J.Ellis, N.E.Mavromatos, C.V.Nanopoulos: {\em Probing Models of
Quantum Space-Time Foam}, gr-qc/9909085;

\bibitem{dopl23} 
R.Haag: {\em Local Quantum Physics}, Texts and Monographs in
Physics, Springer 1994;

\bibitem{dopl24} 
J.Madore, S.Schraml, P.Schupp, J.Wess: {\em Gauge theory on noncommutative
spaces}, hep-th/0001203; B.Jurco, L.Mueller, S.Schraml, P.Schupp, J.Wess:
{\em Construction of non-abelian gauge theories on noncommutative spaces},
hep-th/0104153.

\bibitem{dopl25}
D.Bahns, S.Doplicher, K.Fredenhagen, G. Piacitelli: {\em On the unitarity
problem in space/time noncommutative theories}, Phys. Lett. B553 (2002)
178 - 181, hep-th/0201222.

\bibitem{dopl26}
D.Bahns, S.Doplicher, K.Fredenhagen, G. Piacitelli: {\em Ultraviolet
finite quantum field theory on quantum spacetime}, hep-th/0301100.

\bibitem{dopl27}
D.Bahns, S.Doplicher, K.Fredenhagen, G. Piacitelli: work in progress.

\end{thebibliography}
\end{document}